\documentclass[reprint,superscriptaddress, nofootinbib,aps,prd]{revtex4-1}

\usepackage{amsmath,amsfonts,amssymb}
\usepackage{appendix}
\usepackage{color}
\usepackage{datetime}
\usepackage{graphicx}
\usepackage{dsfont}
\usepackage{array}
\usepackage[citecolor=blue]{hyperref}
\usepackage{braket}

\newcommand{\mL}{\mathcal{L}}
\newcommand{\mO}{\mathcal{O}}

\newcommand{\pd}{\partial}

\begin{document}

\title{The OPE meets semiclassics}
\author{Gabriel Cuomo}
\email{gcuomo@scgp.stonybrook.edu}
\affiliation{Simons Center for Geometry and Physics, Stony Brook University, Stony Brook, NY 11794, USA}
\affiliation{C. N. Yang Institute for Theoretical Physics, Stony Brook University, Stony Brook, NY 11794, USA}
\date{\today}

\begin{abstract}
 \makeatletter\@fileswfalse\makeatother
We show that the correlator of three large charge operators with minimal scaling dimension can be computed semiclassically in CFTs with a $U(1)$ symmetry for arbitrary fixed values of the ratios of their charges. We obtain explicitly the OPE coefficient from the numerical solution of a nonlinear boundary value problem in the conformal superfluid EFT in $3d$. The result applies in all three-dimensional CFTs with a $U(1)$ symmetry whose large charge sector is a superfluid. 
%
%
\end{abstract}
\maketitle

\section{Introduction}

The emergence of classical physics at large quantum numbers is a concept familiar since the early developments of quantum-mechanics \cite{landau1958quantum}. Well-known examples include the rigid rotor and the hydrogen atom. In both these cases, the path integral computing the wave function of states with large angular momentum $\ell\gg 1$ is dominated by a classical saddle-point trajectory. The expansion around this classical configuration coincides with an expansion in inverse powers of $\ell$ (see e.g. \cite{MoninCFT} for an illustration). A similar picture underlies certain simplifications in the study quantum field theories, such as the BMN limit in $\mathcal{N}=4$ super-Yang-Mills theory \cite{Berenstein:2002jq}, the large spin limit of double-trace operators in conformal theories (CFTs) \cite{Alday:2007mf,Komargodski:2012ek,Fitzpatrick:2012yx} or the effective description of large spin mesons as semiclassical rotating strings \cite{Hellerman:2013kba}. Recently, the connection between semiclassics and large quantum number has also provided a useful inspiration in the study of operators carrying a large conserved internal charge in conformal field theories. 

The interest toward this problem stems from the remarkable fact that the large charge sector of many, otherwise nearly intractable, strongly coupled CFTs has been argued to admit a \emph{universal} weakly coupled description in terms of a certain number of \emph{hydrodynamic} Goldstone modes \cite{Hellerman}. This is because, by the state-operator correspondence, large charge operators are associated with states with large charge density for the theory quantized on the cylinder. Analogously to the aforementioned quantum-mechanical examples, one then expects the path integral describing their time evolution to be dominated by a classical trajectory. This in turn will generically break the internal group as well as certain spacetime symmetries \cite{MoninCFT}. Taking the internal group to be $G=U(1)$ for concreteness, the simplest possibility for the symmetry-breaking pattern is the one defining a superfluid phase, which admits a simple effective field theory (EFT) description in terms of a single Goldstone mode \cite{Son:2002zn}. In the EFT, the systematic derivative expansion coincides with an expansion in inverse powers of the charge; this makes it possible to study perturbatively the spectrum of charged operators of the theory. Extensions of this idea have also been applied in related problems in CFTs, such as the study of $R$-charged operators in superconformal theories \cite{Hellerman:2017veg,Hellerman:2017sur} or the determination of the spectrum of spinning charged operators \cite{Cuomo:2017vzg}.

The cylinder viewpoint is illuminating in the study of the spectrum of charged operators. However, it does not seem to provide a useful starting point in the study of correlators with more than two large charge operator insertions,\footnote{It is trivial to extend the analyses of \cite{Hellerman,Badel} to compute correlation functions of light operators in between two large charge operators; see e.g. \cite{MoninCFT,Jafferis:2017zna} or the discussion below Eq. \eqref{eq_Rho_class}.} that so far eluded our understanding (in the non-supersymmetric case\footnote{For theories with extended supersymmetry a lot of progress has been achieved in the study of protected operators, see e.g. \cite{Grassi:2019txd,Bargheer:2019kxb,Bargheer:2019exp,Hellerman:2020sqj} in a similar context. }). In fact, from that perspective it is not even clear if correlators of $n\geq 3$ large charge operators should be calculable within EFT. This is because known EFT descriptions of CFT operators are often associated with the existence of an appropriate macroscopic limit \cite{Lashkari:2016vgj,Jafferis:2017zna}, in which the radius of the sphere is sent to infinity while the local charge and energy densities are kept finite. This limit is most naturally formulated for correlators involving two heavy operators, hence with large scaling dimension, and possibly additional insertions of light operators, whose scaling dimension is much smaller than the heavy ones and whose insertion can be thought as a small perturbation of the heavy states.

However, as we will argue in this paper, the EFT in fact allows also for the controlled calculation of three- and higher-point function of large charge operators. Physically this is because, in radial quantization, each operator creates a superfluid state with charge $n_a\gg 1$ around the insertion point $x_a$. The three-point function then describes the transition between the different superfluids, which is associated with a new classical trajectory in the path integral. Since the radial fields are locally gapped around each state, the corresponding modes are not excited by the classical profile, that thus can be reliably computed within EFT.

Let us call $\mO_n$ the operator with minimal scaling dimension with $U(1)$ charge $n$ and $\bar{\mO}_n$ its hermitian conjugate. Focusing on three-dimensional CFTs, we will show that the calculation of the three-point function $\langle \mO_{n_1}\mO_{n_2}\bar{\mO}_{n_1+n_2}\rangle$, to leading order in the charge $n_1$ and for arbitrary values of the ratio $n_2/n_1$, reduces to the solution of a boundary value problem. Solving this problem numerically, we obtain the operator product expansion (OPE) coefficient. The result takes the form
\begin{equation}\label{EQ_1}
\lambda_{n_1,n_2,\overline{n_1+n_2}}=\exp\left[\frac{ n_1^{3/2}}{6\pi\sqrt{ c}}
f\left(y\right)\right]\,,\qquad
y^2=\frac{n_2}{n_1}\,.
\end{equation}
Here $c$ is the unique Wilson coefficient upon which the EFT depends at leading order and the function $f(y)$ is plotted in Fig.~\ref{fig1}. We expect that the result Eq. \eqref{EQ_1} applies in several theories. For instance, we expect that Eq. \eqref{EQ_1} describes the OPE coefficient of the three operators with minimal scaling dimension at fixed large values of their charges $n_1\sim n_2\sim|n_1+ n_2|\gg 1$ in the $O(2)$ model, where Monte Carlo simulations found $c\approx 0.31$ \cite{Banerjee:2017fcx}. To the best of our knowledge, Eq. \eqref{EQ_1} represents the only example of a universal correlator of three heavy operators in CFT which has been computed (so far) without supersymmetry in $d>2$.\footnote{For related results in $d=2$ see  \cite{Gross:1987ar} and more recently \cite{Cardy:2017qhl,Collier:2019weq,Belin:2020hea}.}

While the main focus of this paper is on the general superfluid EFT, our ideas might also be of some relevance in the study of weakly coupled theories. It has been recently shown that a semiclassical approach can be used to overcome the breakdown of perturbation theory in the determination of the scaling dimensions of large charge operators in the epsilon expansion \cite{Badel}. In Appendix \ref{Appendix} we discuss the semiclassical problem associated with the calculation of the three-point function $\langle\phi^{n_1}\phi^{n_2}\bar{\phi}^{n_1+n_2}\rangle$ in the $|\phi|^6$ tricritical model in $3-\varepsilon$ dimension. There we show that our formulation allows to overcome some technical issues with the generalization of the approach of \cite{Badel} to higher-point functions. The formulation of the problem will make manifest that the calculation reduces to the one in the EFT discussed in the main text in the appropriate regime. A numerical solution can be obtained in more general regimes as well, but we leave a detailed investigation for future work.

The rest of the paper is organized as follows. In Sec. \ref{SecEFT} we briefly review the structure of the EFT and explain how to use it to compute flat space correlators in the large charge regime. In Sec. \ref{SEC3} we discuss the three-point function. We formulate and solve the associated boundary value problem in Sec. \ref{SecNumerics}. The result is discussed in Sec. \ref{SecAnalysis}. We comment on future directions in Sec. \ref{SecOutlook}. The Appendix contains a formal discussion of three-point functions of large charge operators in the weakly coupled $U(1)$ tricritical model in $3-\varepsilon$ dimensions.

\section{Flat space correlators from EFT}\label{SecEFT}

We expect that correlators of large charge operators in three-dimensional CFTs admit a universal EFT description in terms of a single $U(1)$ Goldstone boson \cite{Hellerman,MoninCFT}:
\begin{equation}\label{eqEFT}
S=-\frac{c}{3}\int d^3x\left[|\pd\chi|^3+O\left(\frac{(\pd^2\chi)^2}{|\pd\chi|}\right)\right]\,.
\end{equation} 
Previous works generically considered the action \eqref{eqEFT} on $\mathbb{R}\times S^2$, which provides a natural starting point to study the spectrum of the theory. However, the Weyl invariance of the action ensures that the same results can be recovered working directly in flat space. As a warmup to the study of the three-point function, here we illustrate this point explicitly for the calculation of the two-point function of the charge $n$ operator with minimal scaling dimension.

In order to proceed, we should find a suitable definition of the operator. To this aim, we notice that, by the state-operator correspondence, an insertion of the operator $\mO_n$ at the point $x_i$ in a generic correlator may be equivalently represented by \emph{cutting} the path integral around a ball $B(x_i,r)$ of radius $r$, and specifying appropriate boundary conditions for the fields on the surface $\partial B(x_i,r)$ \cite{Simmons-Duffin:2016gjk}. These follow from performing the path integral inside the ball. For $r\rightarrow 0$ the path integral inside $B(x_i,r)$ is independent of the other operator insertions; hence in this limit we can use the boundary conditions on $\partial B(x_i,r)$ to fully specify the operator. Notice that the infinitesimal radius $r$ of the ball $B(x_i,r)$ provides a natural regulator for the divergences associated with the wave function renormalization of the operator.

In principle, the appropriate definition of the state on $\pd B(x_i,r)$, and hence of the corresponding operator, can always be found solving the functional Schr\"odinger equation in radial quantization. In practice, for the operator $\mO_n$ it is easier to just guess the result. Here we just provide the prescription. We will see later that this indeed reproduces the expected form of the two-point function, hence proving that our guess is correct.

To specify an insertion of the operator with minimal scaling dimension with charge $n$ at the point $x_i$, we add to the action a boundary term of the form \cite{MoninCFT}
\begin{equation}\label{eq:BTeft}
S_{B}^{(i)}=-i\frac{n}{4\pi}\int_{\partial B(x_i,r)} d\Omega\,\chi\,,
\end{equation}
where $d\Omega$ is the volume element on $S^2$. The boundary term \eqref{eq:BTeft} may be thought as a wave functional for the state in radial quantization and fixes its charge to be $n$. Notice that imposing that the variation of the action \eqref{eqEFT} plus the boundary term \eqref{eq:BTeft} vanishes implies that, on the saddle-point solution, the Noether current approaches the form expected from spherical symmetry and Gauss's law close to the insertion point:
\begin{equation}\label{eq:chiBCeft}
J_\mu=ic\,\pd_\mu\chi|\pd\chi|
\stackrel{x\rightarrow x_i}{\longrightarrow}
\frac{n}{4\pi}\frac{(x-x_i)_\mu}{(x-x_i)^3}\,.
\end{equation}
In general, to compute $n$-point functions of the operator $\mO_n$, we cut the path integral around $n$ different points with balls of infinitesimal radius $r$, and we add a boundary term of the form \eqref{eq:BTeft} for each insertion. 

We are now ready to compute the two-point function. The leading-order result is given by:
\begin{equation}\label{EQ_Master2eft}
\langle\bar{\mO}_n(x_f)\mO_n(x_i)\rangle
\simeq \exp\left\{-\left[S+S_{B}^{(i)}+S_{B}^{(f)}\right]_{classical}\right\}\,,
\end{equation}
where 
the action and the boundary terms are calculated on the profile which solves the saddle-point equation:
\begin{equation}\label{Eq_EFTeom}
\pd_\mu J^\mu=0\,,
\end{equation}
with boundary conditions \eqref{eq:chiBCeft} for $x\rightarrow x_{i/f}$ (with the replacement $n\rightarrow -n$ for $x\rightarrow x_f$). The solution reads:
\begin{equation}\label{eqEFTsol}
\chi=-i\mu\log\frac{|x-x_i|}{|x-x_f|}\,,\qquad
\mu=\sqrt{\frac{n}{4\pi c}}\,.
\end{equation}
The physical meaning of this solution is simply understood. Set $x_i$ to be the origin of space and take the limit $x_f\rightarrow\infty$. Then Eq. \eqref{eqEFTsol} reads:
\begin{equation}\label{eq_sflu}
\chi=-i\mu \tau+\text{const.}\,,\qquad
e^{\tau}=|x-x_i|\,,
\end{equation}
which, unsurprisingly, is nothing but the superfluid profile considered in \cite{Hellerman}. The conformal equivalence with the solution \eqref{eq_sflu} and the Weyl invariance of the theory ensure that higher derivative corrections are always suppressed with respect to the leading order for $\mu\sim\sqrt{n}\gg 1$.

We may now use the solution \eqref{eqEFTsol} to compute the classical action \eqref{EQ_Master2eft}. We find:
\begin{equation}
\langle\bar{\mO}_n(x_f)\mO_n(x_i)\rangle=\frac{|\mathcal{N}_n|^2}{x_{if}^{2\Delta_n}}\,,
\end{equation}
where $x_{if}=|x_i-x_f|$ and 
\begin{equation}\label{eqEFTdim}
|\mathcal{N}_n|^2=r^{2\Delta_n}\,,\qquad\Delta_n=\frac{2 n^{3/2}}{3\sqrt{4\pi c}}+O\left(n^{1/2}\right)
\,.
\end{equation}
This agrees with the result for the scaling dimension first obtained in \cite{Hellerman}. The value of the normalization $\mathcal{N}_n$ is unphysical, but it will be important to properly (re)normalize the three-point function and extract the OPE coefficient.

Notice that the form of the solution \eqref{eqEFTsol} is totally fixed by conformal invariance. This is because the classical expectation value for the current physically represents a (normalized) correlator of three primary operators:
\begin{equation}\label{eq_J_class}
J^\mu_{class.}(x)=\frac{\langle\bar{\mO}_n(x_f)J^\mu(x)\mO_n(x_i)\rangle}{\langle\bar{\mO}_n(x_f)\mO_n(x_i)\rangle}\,,
\end{equation}
which is equivalent to Eq. \eqref{eqEFTsol} after using the result for the correlator on the right-hand side based on the Ward identities and conformal invariance. This provides an alternative proof that Eq. \eqref{eq:BTeft} indeed defines a scalar primary operator.

Finally we remark that a definition analogous to Eq. \eqref{eq:BTeft} can be used to simplify the study of flat space correlators of large charge operators directly in the UV theory, when the latter is weakly coupled. In that case, the definition of the operator should be supplemented by appropriate boundary conditions for the radial components of the fields of the model on the surface of the ball. We show an example in Sec. \ref{Sec2pt} of the Appendix for the tricritical $U(1)$ fixed point in $3-\varepsilon$ dimensions.

\section{The three-point function}\label{SEC3}

In the previous section we explained how to compute flat space correlators and we exemplified our approach with the calculation of the two-point function. In this section we formulate and solve the semiclassical problem associated with the three-point function. Though we were only able to obtain a numerical result, we provide approximate analytic expressions in different limits.

Physically, the reason why we can compute $n$-point functions within the EFT is that all the operator insertions specify a superfluid state with large chemical potential $\mu_a^2\sim  |n_a|/c$. Local excitations of the radial mode are separated by a large gap $\sim |\mu_a|$ with respect to the Goldstone modes close to these points. Hence, if the transition between the different superfluid states is sufficiently smooth, these will not be excited by the solution - which can then be computed within EFT. More technically, the calculation of the saddle-point requires control over the full nonlinear structure of the leading-order action \eqref{eqEFT}, but it only receives small corrections from the higher derivative terms.\footnote{This is analogous to what happens in gravity, where many interesting solutions follow from the nonlinear structure of the Einstein-Hilbert action, but they are largely unaffected by higher curvature corrections.}

\subsection{Numerical solution}\label{SecNumerics}

The general form of the three-point function is:
\begin{multline}\label{Eq_3ptGeneral}
\langle\bar{\mO}_{n_2+n_1}(x_f)\mO_{n_2}(x_c)\mO_{n_1}(x_i)\rangle
\\
=\lambda_{n_1,n_2,\overline{n_1+n_2}}\times\frac{\mathcal{N}_{n_1}\mathcal{N}_{n_2}\overline{\mathcal{N}}_{n_1+n_2}}{x_{ic}^{\Delta_{icf}}
x_{cf}^{\Delta_{cfi}}
x_{fi}^{\Delta_{fic}}
}\,,
\end{multline}
where $\lambda_{n_1,n_2,\overline{n_1+n_2}}$ is the OPE coefficient and we defined
\begin{equation}
\Delta_{abc}=\Delta_a+\Delta_b-\Delta_c\,,
\end{equation}
with $\Delta_i=\Delta_{n_1}$, $\Delta_c=\Delta_{n_2}$ and $\Delta_f=\Delta_{n_1+n_2}$.  Notice that since we can always redefine $\mO_n\rightarrow e^{i\alpha_n}\mO_n$, with $\alpha_n\in\mathbb{R}$, the phase of the OPE coefficient is unphysical. For concreteness we consider $n_1\geq n_2\geq 0$.

According to the prescription explained in Sec. \ref{SecEFT}, to compute the three-point function we need to solve the problem specified by the action:
\begin{equation}\label{eqAction3pt}
S_{eff}=S+S_B^{(i)}+S_B^{(c)}+S_B^{(f)}\,,
\end{equation}
where 
\begin{equation}
S_B^{(a)}=-i\frac{n_a}{4\pi}\int_{\pd B(x_a,r)}d\Omega\chi\,,\quad
a=i,c,f\,,
\end{equation}
with $n_i=n_1$, $n_c=n_2$ and $n_f=-n_1-n_2$. This is equivalent to solving the equation of motion \eqref{Eq_EFTeom} with boundary conditions in the form \eqref{eq:chiBCeft} close to the insertion points.

To write a convenient ansatz for the solution, we notice that the solution depends on four points ($x$ and the insertion points). Conformal invariance then suggests to introduce cross-ratios as
\begin{align}\label{eqUandV}
&u=\frac{(x-x_i)^2x_{cf}^2}{x_{ic}^2(x-x_f)^2}=e^{2\tau}\,,\\\
&v=\frac{(x-x_c)^2x_{if}^2}{x_{ic}^2(x-x_f)^2}=1+e^{2\tau}-2e^{\tau}\cos\theta\,,
\end{align}
where $(\tau,\theta)\in (-\infty,\infty)\times [0,\pi]$.
The solution may then be parametrized as:
\begin{equation}\label{eq_chi_tilde}
\chi(x)=-i\mu\tilde{\chi}(\tau,\theta)\,,\qquad
\mu=\sqrt{\frac{n_1}{4\pi c}}\,,
\end{equation}
where $\tilde{\chi}$ is purely real and $\tau$ and $\theta$ are defined in Eq. \eqref{eqUandV}. Formally, this ansatz can be justified by arguments similar to that explained above Eq. \eqref{eq_J_class}. The physical meaning of the coordinates \eqref{eqUandV} is understood noticing that the bulk equation of motion \eqref{Eq_EFTeom} is equivalent to current conservation for the theory on $\mathbb{R}\times S^{2}$ in terms of $\tau$ and $\theta$.

\begin{figure}[t]
\centering
\includegraphics[scale=0.16,trim= 3cm 0cm 3cm 3cm]{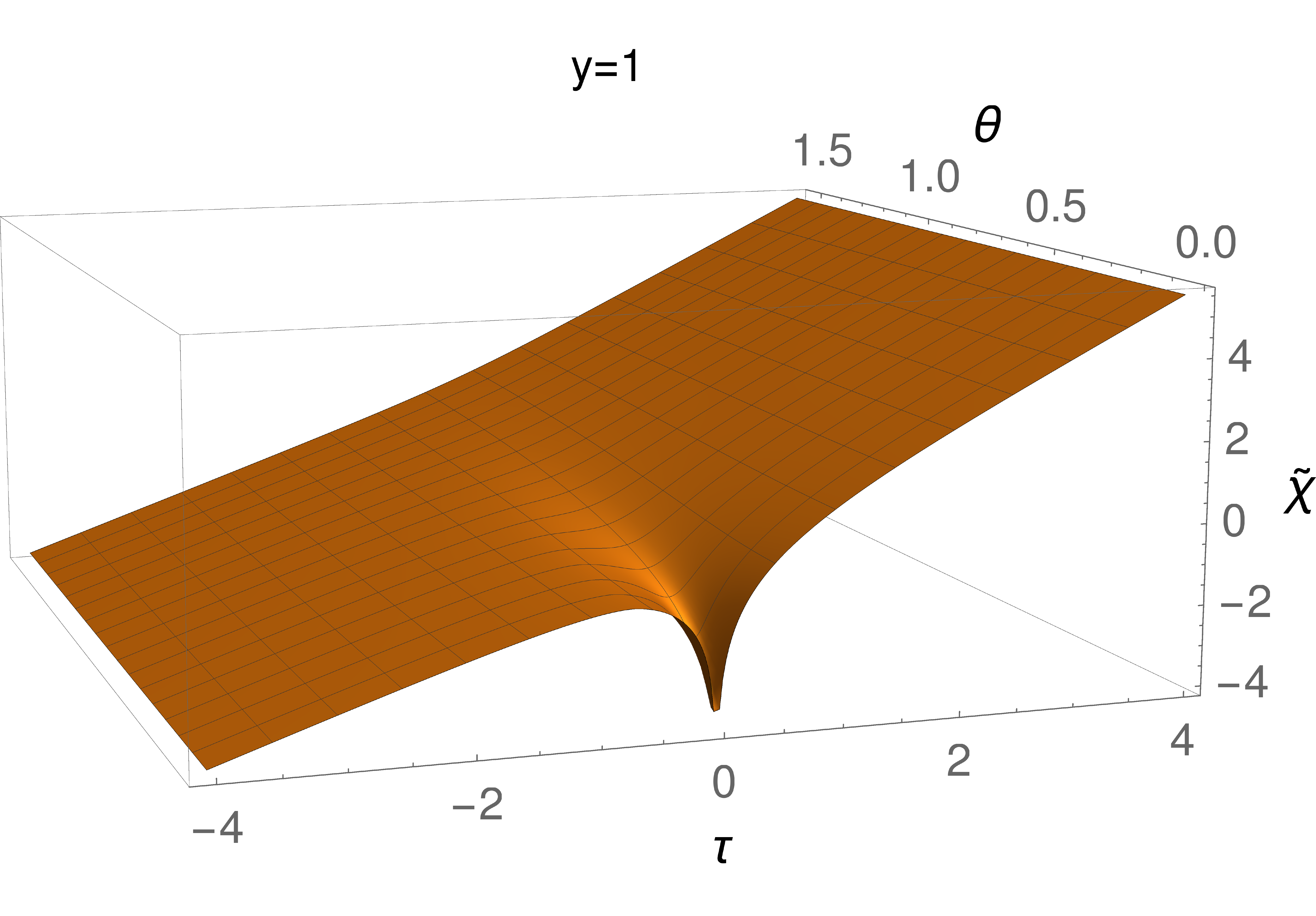}
\caption{Plot of the solution $\tilde{\chi}(\tau,\theta)$ when $y=\sqrt{n_2/n_1}=1$ for $(\tau,\theta)\in[-4,4]\times[0,\pi/2]$. }\label{figP}
\end{figure}
Working with the rescaled field $\tilde{\chi}$ in Eq. \eqref{eq_chi_tilde} is convenient since its value depends on the ratio $y\equiv\sqrt{n_2/n_1}$, but not on $\mu\sim \sqrt{n_1}$. Indeed, the bulk Eq. \eqref{Eq_EFTeom} is independent of any field rescaling, while the boundary conditions for $\tilde{\chi}$ depend only on $y$. Explicitly, the expansion of the field at the boundary points read:
\begin{align}\nonumber
&\tilde{\chi}\xrightarrow{\tau\rightarrow-\infty} \tau+\tilde{c}_i+O\left(e^{\tau}\right)\,,\\
&\tilde{\chi}\xrightarrow[\theta\rightarrow 0]{\tau\rightarrow 0}\frac{y}{2}\log v
+\tilde{c}_c+O\left(\sqrt{v}\right)  \label{eqBoundaryEFT}
\,,\\
&\tilde{\chi}\xrightarrow{\tau\rightarrow+\infty} \sqrt{1+y^2}\,\tau +\tilde{c}_f+O\left(e^{-\tau}\right)\,. \nonumber
\end{align}
Eq. \eqref{eqBoundaryEFT} shows that, in this parametrization, the saddle-point profile takes a superfluid form at $\tau\rightarrow\pm\infty$. The singularity at $(\tau,\theta)=(0,0)$ provides the source which accounts for the mismatch in the charge fluxes at $\tau=-\infty $ and $\tau=\infty$. In Eq. \eqref{eqBoundaryEFT}, besides showing the leading behavior to be imposed in the solution, we also defined the first constant corrections $\tilde{c}_{i/c/f}$ to the field $\chi$ close to the insertion points, whose value is determined solving the equations of motion. Finally we stressed that the additional corrections to the boundary values at $\tau\rightarrow\pm\infty$ decay exponentially fast.\footnote{The form of the solution close to the insertion points is obtained linearizing the equation around the profile dictated by the boundary conditions.} 

Discretizing the equation as explained in \cite{Krikun:2018ufr} and introducing a sufficiently large cutoff for $|\tau|\leq T$, it is now possible to solve for the profile numerically. The typical shape of the solution is shown in Fig. \ref{figP}. 
As it may be seen from there, $\tilde{\chi}(\tau,\theta)$ is regular everywhere but at the insertion points, where the solution however is conformally equivalent to a superfluid profile. Hence higher derivative corrections to the EFT \eqref{eqEFT} are always suppressed by $1/\mu^2$ with respect to the leading term, provided $ y\gg 1/\mu$. This justifies the use of the action \eqref{eqEFT} \emph{a posteriori}.

\begin{figure}[t]
\centering
\includegraphics[scale=0.177]{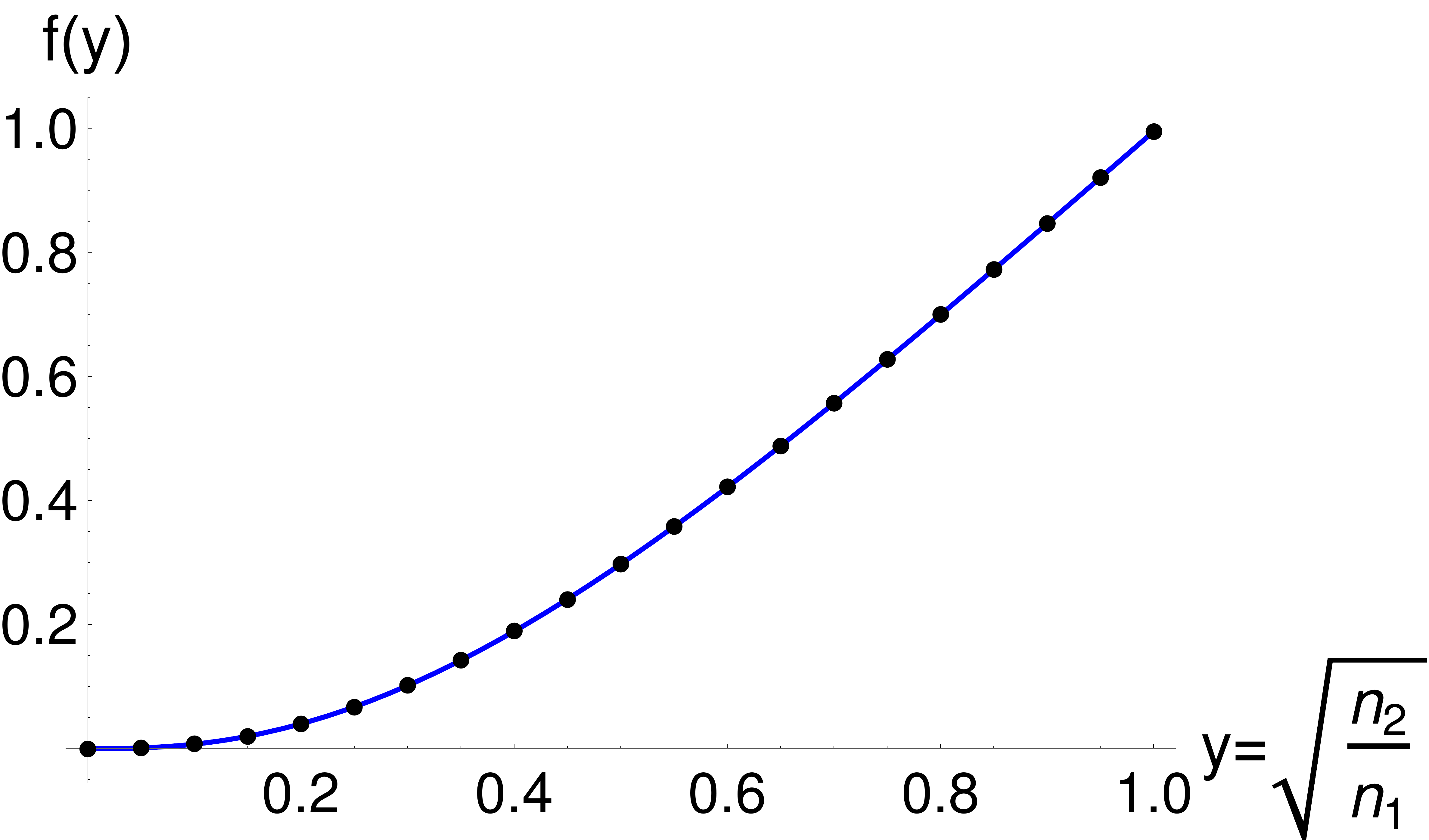}
\caption{The function $f(y)$ determining the OPE coefficient \eqref{eqOPEfinal}. The blue line interpolates between the numerical values.}\label{fig1}
\end{figure}
We now use the numerical solution for the profile to extract the OPE coefficient. To this aim we notice that the action \eqref{eqAction3pt} on shell reduces to a boundary term. Therefore, we find that the OPE coefficient may be written purely in terms of the boundary coefficients $\tilde{c}_a$ defined in eq. \eqref{eqBoundaryEFT}:
\begin{equation}\label{eqOPEfinal}
\begin{split}
\lambda_{n_1,n_2,\overline{n_1+n_2}}&=\exp\left\{\frac{2}{3}n_1\mu\left[\tilde{c}_i+y^2\tilde{c}_c-(1+y^2)\tilde{c}_f\right]\right\}\\
&=\exp\left[\frac{ n_1^{3/2}}{6\pi\sqrt{ c}}f(y)\right]\,,
\end{split}
\end{equation}
where $f(y)=\sqrt{4\pi}\left[\tilde{c}_i+y^2\tilde{c}_c-(1+y^2)\tilde{c}_f\right]$. From the behavior of the numerical solution close to the insertion points we determine the $\tilde{c}_a$'s and we use them to compute the function $f(y)$.
The numerical result is plotted in Fig. \ref{fig1} for $0\leq y\leq 1$. This range fully determines the OPE coefficient, since by permutation symmetry we must have $f(y)=y^3f(1/y)$. We verified that this relation is satisfied within the numerical error for $1\leq y\leq 2$. \footnote{The result plotted in \ref{fig1} was obtained discretizing the equation on a grid $[-T,T]\times[0,\pi]$ with $T=8$, uniformly discretized in $651\times 131$ points. Different choices of the grid provided compatible results within the numerical uncertainty. A fourth-order approximation was used to discretize the derivatives. We also made use of the shift symmetry of $\tilde{\chi}$ to set $\tilde{c}_c=0$ in Eq. \eqref{eqBoundaryEFT}. From the variance around the mean value of the field at $\tau=\pm T$, we estimated the numerical error on $\tilde{c}_{i/f}$, from which we found the relative uncertainty on $f(y)$ to be $\lesssim 1\%$ for all values of $y$.}

\subsection{Analysis of the result}\label{SecAnalysis}

Eq. \eqref{eqOPEfinal} with the numerical result plotted in Fig.~\ref{fig1} is the main result of this paper. 
Here we discus its behaviour for small $y\ll 1$ and for $y\simeq 1$.

\begin{figure}[t]
\centering
\includegraphics[scale=0.167, trim= 1.6cm 0cm 0cm 0cm]{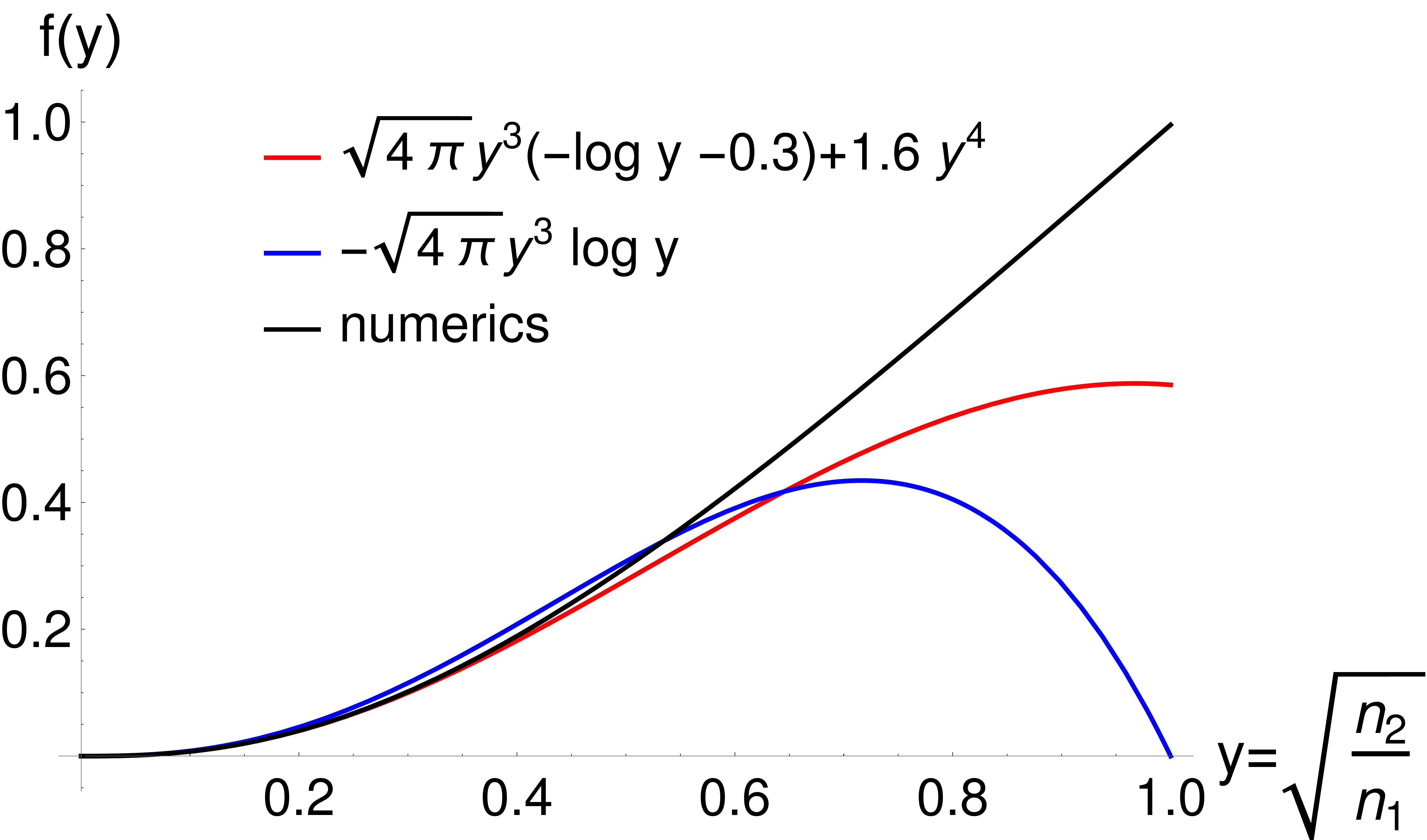}
\caption{Comparison between the numerical result for the function $f(y)$ and the form predicted by Eq. \eqref{eq_small_y} for small $y$.}\label{fig3}
\end{figure}

In the regime $1\ll n_2/c\ll n_1/c$ we may compare the result with the EFT prediction for the OPE coefficient of a light operator in between two large charge ones. This is given by \cite{MoninCFT,Cuomo:2020rgt}:
\begin{multline}
\lambda_{n_1,n_2,\overline{n_1+n_2}}=C_{n_2}\left(\frac{n_1}{c}\right)^{\Delta_{n_2}/2}\\
\times\exp\left[0.05\times\frac{n_2^2}{\sqrt{n_1c/3}}
+O\left(\frac{n_2^{5/2}}{n_1\sqrt{c}}\right)\right]\,,
\end{multline}
where $C_{n_2}$ is a Wilson coefficient, which may depend on $n_2$ but not on 
$n_1$. Such form is compatible with our result \eqref{eqOPEfinal} for $1\ll n_2/c\ll n_1/c$ provided the coefficient $C_{n_2}$ takes the form:
\begin{equation}
C_{n_2}=\left(\frac{c \,e^{2\alpha}}{n_2}\right)^{\Delta_{n_2}/2}\quad\text{for}\quad n_2/c\gg 1\,,
\end{equation}
where $\alpha$ is a constant which does not depend on the charge. This implies that the function $f(y)$ in Eq. \eqref{eqOPEfinal} admits the following expansion for small $y$:
\begin{equation}\label{eq_small_y}
f(y)=\sqrt{4\pi}y^3\left(-\log y+\alpha\right)+1.6\, y^4+O\left(y^5\right)\,.
\end{equation}
This is in good agreement with our numerical result. 
More precisely, retaining only the leading order $f(y)\approx-\sqrt{4\pi}y^3\log y$ we find agreement up to relative $O(10\%)$ for $y\lesssim 0.4$. The comparison of Eq. \eqref{eq_small_y} with our numerical result for $y\lesssim 0.3$ can be used to determine $\alpha\approx -0.3$. Even including the subleading terms, for $y\gtrsim 0.5$, which is equivalent to $n_2/n_1\gtrsim 1/4$, the prediction \eqref{eq_small_y} and the numerical results start to differ significantly. In particular, the expansion \eqref{eq_small_y} obviously cannot be extrapolated to $y=1$. The comparison is summarized in Fig. \ref{fig3}.

For $y\sim O(1)$ the result \eqref{eqOPEfinal} can be seen as a resummation of infinite $n_2/n_1$ perturbative corrections to the OPE coefficient computed in \cite{MoninCFT,Cuomo:2020rgt} for $n_2\ll n_1$. We find $f(1)=0.996\approx 1$ and in general $f(y)\sim O(1)$ for $y=\sqrt{n_2/n_1}\sim O(1)$. Hence Eq. \eqref{eqOPEfinal} predicts that the fusion coefficient for three operators of charge $n_a\sim n\gg 1$ grows faster than exponentially, as $\sim e^{\# n^{3/2}/\sqrt{c}}\sim e^{\# \Delta_n}$. As for the scaling dimension \eqref{eqEFTdim}, the power $n^{3/2}$ just follows from dimensional analysis and semiclassics. In fact the value of the Lagrangian, locally, is naturally set by the largest dimensionful scale for the theory on $\mathbb{R}\times S^2$, which is given by the charge density $J_0\sim n$. Dimensional analysis implies $\mL\sim J_0^{3/2}$, from which the scaling follows. Notice also that the non-polynomial growth in Eq. \eqref{eqOPEfinal} is not compatible with a finite macroscopic limit as defined in \cite{Jafferis:2017zna}.

Eq. \eqref{eqOPEfinal} is reminiscent of the result of \cite{Cardy:2017qhl,Collier:2019weq}, where it was shown that the average value of the OPE coefficient for three heavy operators obtained in $2d$ CFTs grows as $c^\Delta$, where $c>1$ is an $O(1)$ number. As commented in \cite{Cardy:2017qhl}, this growth is  not surprising. Indeed Eq. \eqref{eqOPEfinal} describes a correlation function with operators located at $0$, $1$ and $\infty$, but this is only a convention. A different choice of the operator positions would change the correlation function by an exponential factor, and could possibly cancel the growth of the result \eqref{eqOPEfinal}. 

We may also use permutation symmetry $f(y)=y^3f(1/y)$ to provide an approximate expression for $f(y)$ when $y\simeq 1$. Indeed Taylor expanding the relation $f(y)=y^3f(1/y)$ we conclude $f'(1)=\frac{3}{2}f(1)$, from which we infer:\footnote{In general, the functional equation $f(y)=y^3f(1/y)$ relates $f^{(2n+1)}(1)$ and a linear combination of even derivatives of lower order, $f^{(2n) }(1),\,f^{(2n-2)}(1),\ldots,f(1)$. For instance, the value of $f''(1)$ can be used to improve the approximation \eqref{eq_approx_1} with two more terms. Comparing different numerical fits we found a relatively small value for the second derivative $ f''(1)\lesssim 0.08$, justifying the accuracy of the linear interpolation \eqref{eq_approx_1}.}
\begin{equation}\label{eq_approx_1}
\begin{split}
f(y)&\simeq f(1)\left[1-\frac{3}{2}(1-y)+O\left((1-y)^2\right)\right]\\
&\simeq 1-\frac32(1-y)+O\left((1-y)^2\right)\,.
\end{split}
\end{equation}
Eq. \eqref{eq_approx_1} agrees with our numerical result up to relative $O\left(10\%\right)$ for $y\gtrsim 0.55$.

\begin{figure}[t]
\centering
\includegraphics[scale=0.17, trim= 1.6cm 0cm 0cm 0cm]{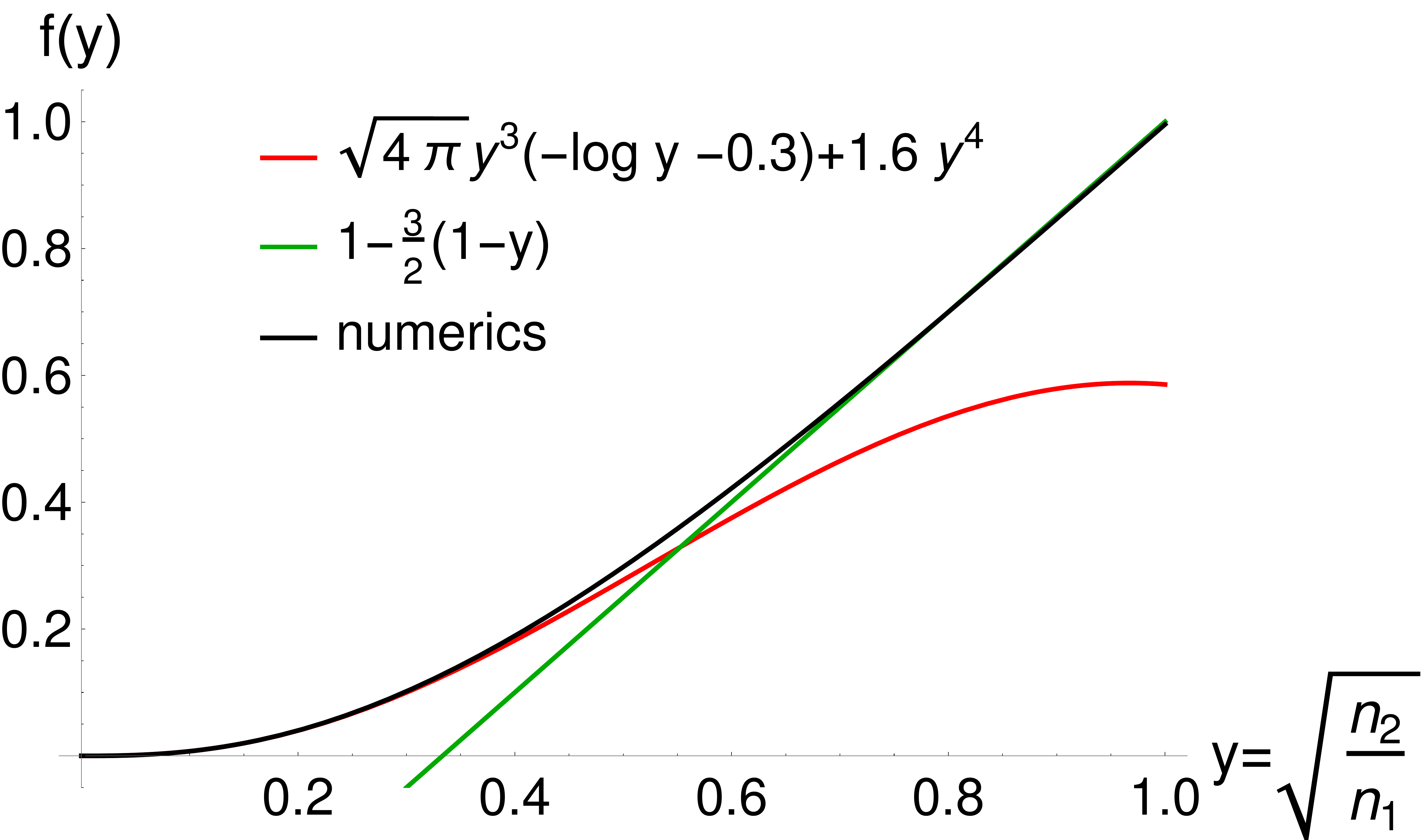}
\caption{Comparison between the numerical result for the function $f(y)$ and the approximate expansions \eqref{eq_small_y} and \eqref{eq_approx_1}.}\label{fig4}
\end{figure}

We summarize the comparison between the approximations that we discussed and the numerical result in Fig. \ref{fig4}. We see that, though we have not been able to compute the OPE coefficient \eqref{eqOPEfinal} in a closed analytic form, the approximate expressions \eqref{eq_small_y} and \eqref{eq_approx_1} combined suffice to give a numerically accurate description of the result up to relative $O(10\%)$ for all values of $y$. 

\section{Outlook}\label{SecOutlook}

In this paper we illustrated how to compute semiclassically three-point functions of large charge operator in three-dimensional $U(1)$ invariant CFTs. Our main result, Eq. \eqref{eqOPEfinal}, provides a novel general prediction for the OPE coefficient of three large charge operators with minimal scaling dimensions. This prediction applies in all CFTs whose large charge sector is in a superfluid phase. For instance, we expect that it applies in the $O(2)$ model, where in principle it may be tested via Monte Carlo simulations, along the lines of \cite{Banerjee:2017fcx}.

Our work admits some obvious extensions. As already mentioned, it is possible to obtain numerical results for the OPE coefficients directly in weakly coupled models, such as the tricritical model in $3-\varepsilon$ dimensions discussed in Appendix \ref{Appendix}, or, via an obvious generalization, in the Wilson-Fisher $O(N)$ fixed-point in $4-\varepsilon$ dimensions. Perhaps, a similar procedure might be applied in the $O(N)$ models at large $N$ as well (see \cite{Alvarez-Gaume:2019biu,Giombi:2020enj} for recent progress in the study of large charge operators at large $N$). It is also possible to study higher-point functions. For instance, generalizing Eq. \eqref{eq_chi_tilde}, we expect that the semiclassical solution $\chi(x)$ determining the correlator of four charged operators will depend on five independent cross ratios. Three of them will be given by the independent components of the coordinate $x$, while the other two should coincide with the two physical cross-ratios specifying the configuration in the four-point function. A detailed analysis of this problem within the large charge EFT would allow for the determination of OPE coefficients involving operators corresponding to \emph{phonon} excitations on top of the superfluid ground state \cite{Hellerman}. It would also be interesting, but probably technically challenging, to extend our analysis to subleading order in derivatives and to include quantum effects. Notice however that in the EFT quantum corrections to the OPE coefficient scale only as $n^0$, and are hence quite suppressed with respect to the leading order. 

To the best of our knowledge, the result \eqref{eqOPEfinal} represents the only example of a universal correlator of three heavy operators in CFT which has been computed without supersymmetry in $d>2$. The universal nature of the result stems from the fact that it has been obtained within the conformal superfluid EFT. Differently from matrix elements of the form heavy-light-heavy, which are obtained considering small perturbations of the heavy state \cite{MoninCFT}, the result was obtained considering a new saddle point profile, which accounts for the classical \emph{nonlinearities} of the action exactly. 
Perhaps a similar approach might extend the existing result on heavy-light-heavy OPE coefficients of generic (not necessarily charged) operators, which are based on thermodynamics and hydrodynamics \cite{Lashkari:2016vgj,Delacretaz:2020nit}, to the heavy-heavy-heavy regime. 

In the case of weakly coupled theories, the semiclassical approach can be used to overcome the breakdown of perturbation theory for multi-legged amplitudes describing large charge operators \cite{Badel}. For instance, in the tricritical $|\phi|^6$ model in $3-\varepsilon$ dimensions discussed in the Appendix \ref{Appendix}, the result \eqref{eqOPEfinal} can be seen as a resummation of infinite Feynman diagrams in the regime $\lambda n_1,\,\lambda n_2\gg 1$, where $\lambda\sim \sqrt{\varepsilon}$ is the perturbative coupling. The connection between semiclassics and multi-legged amplitudes was first appreciated in the study of multiparticle production processes \cite{Rubakov:1995hq}, where the physical rate was shown to be associated with the solution of a semiclassical boundary value problem \cite{Son:1995wz}. Unfortunately, the status on the solution of the latter is still a controversial subject to this date (see e.g. \cite{Khoze:2018mey} and \cite{Monin:2018cbi} for the two opposite viewpoints). It is interesting to compare that semiclassical problem with the one considered in this work.
Somewhat similarly to the situation considered in Sec. \ref{SecNumerics}, where the singularity of the solution in Fig. \ref{figP} ensures that Gauss's law holds globally, the solution to the problem of \cite{Son:1995wz} requires a singular field configuration to account for the change in the energy (the production by a source) between the initial and the final state. However, differently from the case of CFT correlators, the multiparticle production classical problem does not admit a fully Euclidean formulation. Perhaps more importantly, at present there is no clear \emph{EFT picture} for the corresponding physical process. This is again different from the problem that one encounters in the study of large charge CFT operators in perturbative theories, addressed for the first time in \cite{Badel}, whose solution was indeed almost completely guided by previous EFT developments in the same context \cite{Hellerman,MoninCFT}.

\section*{Acknowledgements}

I thank Zohar Komargodski, Alexander Monin, Sridip Pal and Riccardo Rattazzi for inspiring discussions, and I am especially grateful to Jo\~ao Penedones for collaboration at the early stages of this work. It is also a pleasure to thank my father, Massimo Cuomo, for delightful discussions on numerical methods. My work is supported by the Simons Foundation (Simons Collaboration on the Non-perturbative Bootstrap) grants 488647 and 397411.

\appendix

\section{The three-point function boundary value problem in a weakly coupled model}\label{Appendix}

In this Appendix we generalize the approach in the main text to the calculation of three-point functions of large charge operators in weakly coupled theories, focusing on the tricritical $U(1)$ fixed point in $3-\varepsilon$ dimensions for illustration. 

It has been shown in \cite{Badel,Badel2} that the scaling dimension of the operator $\phi^n$ can be computed semiclassically as an expansion in the coupling $\lambda\sim\varepsilon^{1/2}$  at arbitrary fixed values of $\lambda n$, which plays a role analogous to the `t Hooft coupling in large N gauge theories. The same approach has been successfully applied in the calculation of two-point function of charged operators in several different models (see e.g. \cite{Antipin:2020abu,Giombi:2020enj}), but technical difficulties make the generalization of the method to higher-point functions non-straightforward. Here we show that the calculation of the three-point function $\langle\phi^{n_1}\phi^{n_2}\bar{\phi}^{n_1+n_2}\rangle$ in the perturbative model can be reduced, to leading order in $\lambda$ but to all orders in $\lambda n_1$ and $\lambda n_2$, to the solution of a boundary value problem. This problem enjoys exact conformal symmetry and, crucially, it is formulated in three Euclidean dimensions; this makes it always possible to look for a solution numerically, differently from the existing formulation of the problem based on dimensional regularization. We argue that for $\lambda n_1\geq \lambda n_2\gg 1$ the solution can be found within the EFT studied in the main text. We leave for future work a detailed investigation of more general regimes.

\subsection{The model}\label{secModel}

We consider the following (Euclidean) action in $d=3-\varepsilon$ dimensions for a complex scalar field $\phi$:
\begin{equation}\label{eq:action}
S=\frac{1}{\lambda}\int d^dx\left[\pd\bar{\phi}\pd\phi
+\frac{1}{36}(\bar\phi\phi)^3\right]\,,
\end{equation}
The beta function of the coupling starts at two-loop order, hence this model is conformal in $d=3$ up to $O(\lambda)$. To higher orders, the theory admits an IR stable fixed point in $3-\varepsilon$ dimensions with $\varepsilon\ll 1$ for \cite{Pisarski}:
\begin{equation}\label{lambdaStar}
\frac{\lambda_*^2}{(4\pi)^2}=\frac{3 }{7}\varepsilon+O(\varepsilon^2).
\end{equation}
We will always work at leading order in the coupling (in the large charge double scaling limit); hence in what follows we set $d=3$ when not specified otherwise.

\subsection{Operator insertions as boundary conditions in the double-scaling limit}\label{SecOpDefWC}

We consider correlation functions of the lowest dimensional operator of charge $n$, given by $\phi^n$ within perturbation theory. In \cite{Badel} it was shown that its two-point function can be computed in the double-scaling limit $\lambda\rightarrow 0$, $n\rightarrow\infty$ with $\lambda n$ fixed. The result takes the form
\begin{equation}\label{eq2pt}
\begin{split}
\langle\bar{\phi}^n(x_f)\phi^n(x_i)\rangle
&=n!e^{\frac{1}{\lambda}\Gamma_{-1}(\lambda n,x_{fi})+
\Gamma_0(\lambda n,x_{fi})+\ldots}
\\
&=
\frac{|\mathcal{N}_n|^2}{x_{if}^{2\Delta_n}}\,,
\end{split}
\end{equation}
where $x_{if}=|x_i-x_f|$, $\mathcal{N}_n$ is the normalization of the operator, including its divergent wave function, and $\Delta_n$ is the scaling dimension. In particular, Eq. \eqref{eq2pt} implies that the scaling dimension admits the following expansion:
\begin{equation}
\Delta_n=\frac{1}{\lambda}f_{-1}(\lambda n)+f_0(\lambda n)+\ldots\,,
\end{equation}
The leading-order term $f_{-1}$ arises as the result of a nontrivial saddle point which accounts for the operator insertions, and hence is purely semiclassical in nature. Formally, the saddle point equation follows from exponentiating the operator insertions and treating them as sources in the action. This leads to the following partial differential equations in $d=3-\varepsilon$:
\begin{align}\nonumber
&\left[-\pd^2+\frac{1}{12}\left(\bar{\phi}\phi\right)^2\right]\phi=\frac{\lambda n}{\bar{\phi}(x_f)}\delta^{(d)}(x-x_f)\,,\\  \label{eq_saddle_dimreg}
&\left[-\pd^2+\frac{1}{12}\left(\bar{\phi}\phi\right)^2\right]\bar\phi=\frac{\lambda n}{\phi(x_i)}\delta^{(d)}(x-x_f)\,.
\end{align}
For small values of $\lambda n$ we can make sense of these equations working within dimensional regularization and treating the interaction term perturbatively (see e.g. \cite{Badel,Arias-Tamargo:2019kfr}). Unfortunately, it is less clear how to proceed for more general values of $\lambda n$. To appreciate this, notice that the leading-order term $\Gamma_{-1}$ in Eq. \eqref{eq2pt}, while it is not affected by the renormalization of the coupling, it does contain the divergences associated with the wave function renormalization of the operator.
This implies that the Eqs. \eqref{eq_saddle_dimreg} must be solved via analytic continuation to $d=3-\varepsilon$ dimensions to regularize the result. This procedure breaks scale invariance at intermediate steps of the calculation, and it does not allow for a simple numerical formulation of the problem.\footnote{The authors of \cite{Giombi:2020enj} were able to solve analytically a similar saddle point problem to determine the two-point function of large traceless symmetric operators in the $O(N)$ model at large $N$. A similar approach might allow to solve Eqs. \eqref{eq_saddle_dimreg} as well, but we do not expect it to be easily generalizable to higher-point functions, in which case the saddle point solution might not be expressible in a closed analytic form.} 

In \cite{Badel} this issue was solved exploiting the Weyl invariance of the model to map the theory to the cylinder $\mathbb{R}\times S^{d-1}$. The state-operator correspondence then maps the operator $\phi^n$ to the lowest energy state for the theory quantized on $\mathbb{R}\times S^{d-1}$.\footnote{This is because all the other charge $n$ operators have scaling dimension $O(1)$ larger than that of $\phi^n$, and hence cannot mix with it within perturbation theory. Notice that this argument holds for arbitrary values of $\lambda n$, as it follows from the result of \cite{Badel2} for the Fock spectrum of charge $n$ states on the cylinder.
} This allows to extract the scaling dimension $\Delta_{\phi^n}$ from the expectation of the Euclidean evolution operator $e^{-HT}$ in an arbitrary state of charge $n$, provided this has nonzero overlap with the true ground state. A clever choice of the state then allows to perform the calculation straightforwardly. 

While the approach of \cite{Badel} cannot be generalized to higher-point functions, we may proceed as we did for the EFT in Sec. \ref{SecEFT}. Namely, we define the insertion of the operator $\phi^n$ at the point $x_i$ cutting the path integral around a ball $B(x_i,r)$, with $r\rightarrow 0$, and specifying boundary conditions on the surface $\pd B(x_i,r)$.\footnote{More precisely, this procedure specifies an operator which is equal to $\phi^n$ up to an unphysical normalization, which cancels out from observable quantities. With a slight abuse of notation, we shall call also this operator $\phi^n$ in what follows.} Since the infinitesimal radius $r$ of the ball $B(x_i,r)$ provides a natural regulator for the divergences associated with the wave function renormalization, we can work exactly in $d=3$ to leading order.

To choose the appropriate boundary conditions, we map to the plane the state chosen in \cite{Badel2} to perform the calculation on $\mathbb{R}\times S^2$. To this aim it is convenient to work in polar field coordinates:
\begin{equation}\label{eq_polar_coordinates}
\phi=\frac{\rho}{\sqrt{2}}e^{i\chi}\,,\qquad
\bar{\phi}=\frac{\rho}{\sqrt{2}}e^{-i\chi}\,.
\end{equation}
To specify an insertion of the operator with minimal scaling dimension with charge $n$ at the point $x_i$, 
we add to the action a boundary term analogous to Eq. \eqref{eq:BTeft},
\begin{equation}\label{eq:BT1}
S_{B}^{(i)}=-i\frac{n}{4\pi}\int_{\partial B(x_i,r)} d\Omega\,\chi\,.
\end{equation}
As in the case of the EFT, imposing that the variation of the action \eqref{eq:action} plus the boundary term \eqref{eq:BT1} vanishes implies that the Noether current takes the following form close to $x_i$:
\begin{equation}\label{eq:chiBC}
J_\mu=\frac{i}{\lambda}(\pd_\mu\chi)\rho^2
\stackrel{x\rightarrow x_i}{\longrightarrow}
\frac{n}{4\pi}\frac{(x-x_i)_\mu}{(x-x_i)^3}\,.
\end{equation}
To fully specify the boundary value problem, and hence the operator, we need to specify another boundary condition on $\pd B(x_i,r)$. It turns out that the convenient choice is to demand that $\rho$ takes a fixed value, given by
\begin{equation}\label{eq:rhoBC1}
\rho^2\stackrel{x\rightarrow x_i}{\longrightarrow}\frac{f^2}{|x-x_i|}
\qquad\text{with}\qquad \frac{f^4}{48}=\mu^2-\frac14\,,
\end{equation}
where $\mu$ is
\begin{equation}\label{eqMu}
\mu=\text{sgn}(n) \frac{\sqrt{1+\sqrt{1+\frac{(\lambda n)^2}{12 \pi ^2}}}}{2 \sqrt{2}}\,.
\end{equation}
For small $\lambda n\geq 0$ one finds $\mu\simeq\frac12+\frac{(\lambda n)^2}{192 \pi ^2}$, while in the opposite regime $\mu\simeq \frac{1}{4}\left(\frac{\lambda n}{\sqrt{3}\pi}\right)^{1/2}$. Finally, we assume that the path integral is supplemented with vanishing boundary conditions for the fields $\phi$ and $\bar{\phi}$ at infinity. 

In the next section we will compute the classical action obtained by expanding the path integral around the classical solution of the boundary value problem specified by two operator insertions. The result will take indeed the form of a two-point function for a primary operator with scaling dimension $\Delta_{n}$, hence proving that our definition indeed specifies the desired operator to leading order in the coupling. We will then use this definition to discuss the three-point function.

\subsection{The two-point function in flat space}\label{Sec2pt}

We are now ready to compute the two-point function. The leading-order result is given by:
\begin{equation}\label{EQ_Master2}
\begin{split}
\langle\bar{\phi}^n(x_f)\phi^n(x_i)\rangle
&\simeq \exp\left\{-\left[S+S_{B}^{(i)}+S_{B}^{(f)}\right]_{classical}\right\}\,,
\end{split}
\end{equation}
where 
the action and the boundary terms are calculated on the profile which solves the saddle point equation:
\begin{equation}\label{Eq_EOM2}
-\partial^2\phi+\frac{1}{12}\phi(\bar\phi\phi)^2=0\,,\qquad
-\partial^2\bar\phi+\frac{1}{12}\bar\phi(\phi\bar{\phi})^2=0\,,
\end{equation}
with boundary conditions \eqref{eq:chiBC} and \eqref{eq:rhoBC1} for $x\rightarrow x_{i/f}$ (with the replacement $n\rightarrow -n$ for $x\rightarrow x_f$). 

In the limit $r\rightarrow 0$ that we consider the semiclassical problem is conformally invariant. 
It is therefore natural to consider an ansatz in the form of a correlator of three primaries inserted at $x$, $x_i$ and $x_f$ for the classical profile. Therefore we look for solutions of Eqs. \eqref{Eq_EOM2} in the form:
\begin{equation}\label{eqAnsatz2pt}
\begin{split}
\phi&=c\times\langle\mathcal{O}_{\Delta=1/2}(x)
\mathcal{O}_{\Delta=i\eta/2}(x_i)\mathcal{O}_{\Delta=-i\eta/2}(x_f)\rangle\,\\
&=c\times\frac{|x_i-x_f|^{1/2}}{|x-x_i|^{1/2+i\eta}|x-x_f|^{1/2-i\eta}}\,,
\end{split}
\end{equation}
where we used that $\phi(x)$ has weight and mass dimension $1/2$. We use a similar ansatz for $\bar{\phi}$:
\begin{equation}\label{eqAnsatz2pt2}
\begin{split}
\bar{\phi}(x)&=\bar{c}\times\langle\mathcal{O}_{\Delta=1/2}(x)
\mathcal{O}_{\Delta=-i\eta/2}(x_i)\mathcal{O}_{\Delta=+i\eta/2}(x_f)\rangle\,\\
&=\bar{c}\times\frac{|x_i-x_f|^{1/2}}{|x-x_i|^{1/2-i\eta}|x-x_f|^{1/2+i\eta}}\,,
\end{split}
\end{equation}
where we reflected $\eta\rightarrow-\eta$ with respect to Eq. \eqref{eqAnsatz2pt}. Using this ansatz in the equations of motion and demanding the boundary conditions we find:
\begin{align}
\frac{1}{12}(c \bar{c})^2=-\left(\eta^2+\frac{1}{4}\right)\,,\qquad
-2i\eta\, c\bar{c}=\frac{\lambda n}{4\pi}\,.
\end{align}
These equations are solved for purely imaginary $\eta$ as:
\begin{equation}\label{eq_2pt_solution}
\eta=i\mu\,,\qquad\bar{c}c=f^2/2\,,
\end{equation}
where $\mu$ and $f$ are given in Eqs. \eqref{eq:rhoBC1} and \eqref{eqMu}. Therefore the solution is totally specified up to rescalings of the form $\{c,\bar c\}\rightarrow\left\{\alpha c,\alpha^{-1}\bar{c}\right\}$. Indeed, on the solution \eqref{eq_2pt_solution} the fields \eqref{eqAnsatz2pt} and \eqref{eqAnsatz2pt2} are analytically continued away from the contour $\phi=\bar{\phi}^*$, and rescalings of the form $\{\phi,\bar\phi\}\rightarrow\left\{\alpha\phi,\alpha^{-1}\bar{\phi}\right\}$ are a symmetry of the action analytically continued to arbitrary values of the fields.

To understand the physical meaning of this solution, it is convenient to set $x_i$ to be the origin of space and take the limit $x_f\rightarrow\infty$. Then working in polar coordinates \eqref{eq_polar_coordinates} for the field we recast the profile as:
\begin{equation}\label{eq_2pt_cyl}
\rho^2_{cyl}=|x|\rho^2=f^2\,,\qquad
\chi=-i\mu\log|x|+\text{const.}\,.
\end{equation}
Introducing a time coordinate $e^{\tau}=\log|x|$ and performing a Weyl map to the cylinder, Eq. \eqref{eq_2pt_cyl} coincides with the superfluid profile considered in \cite{Badel2} for the theory on $\mathbb{R}\times S^2$. Of course, this is not surprising, since our definition of the operator coincides with the prescription used to specify the charged state in \cite{Badel2}. 

We may now find the result for the two-point function evaluating the action on the corresponding solution. Upon integrating by parts, we find:
\begin{multline}\label{eq_2pt_ClassicalAction}
\left[S+S_{B}^{(i)}+S_{B}^{(f)}\right]_{classical}\\=-\frac{1}{\lambda}
\left(\lambda n \mu \log \frac{r^2}{x_{if}^2}-2\pi f^2+\frac{1}{144}
\int_{\mathbb{R}'^3}d^3x\rho^6\right)\,,
\end{multline}
where the bulk integral is over the subtracted space $\mathbb{R}'^3=\mathbb{R}^3\setminus\left[ B(x_i,r)\cup B(x_f,r)\right]$. The explicit evaluation gives:
\begin{equation}\label{eq_2pt_Integral}
\int_{\mathbb{R}'^3}d^3x\rho^6=4\pi f^6
\log\frac{x_{if}^2}{r^2}\,.
\end{equation}
Overall, using Eqs. \eqref{eq_2pt_ClassicalAction} and \eqref{eq_2pt_Integral} in Eq. \eqref{EQ_Master2}, we find that the two-point function takes the form \eqref{eq2pt} expected for the primary operator $\phi^n$. The leading-order results in the double-scaling limit \eqref{eq2pt} for the operator normalization and the scaling dimension read:
\begin{equation}\label{eq2ptResultPre}
|\mathcal{N}_n|^2=r^{2\Delta_n}e^{-2\pi f^2/\lambda}\,,
\end{equation}
\begin{equation}\label{eq2ptResult}
\begin{split}
\Delta_n&=\frac{n}{3}\left(2\mu+\frac{1}{4\mu}\right)\\
&=
\begin{cases}
\frac{n}{2}+\frac{\lambda^2 n^3}{576 \pi ^2}+\ldots &\text{for } \lambda n\ll 4\pi\,,\\
\frac{\sqrt{\lambda}n^{3/2}}{3^{1/4}6  \sqrt{\pi }}+\ldots &\text{for } \lambda n\gg 4\pi\,.
\end{cases}
\end{split}
\end{equation}
The result \eqref{eq2ptResult} for the scaling dimension agrees with the one found in \cite{Badel2}, hence proving that our prescription indeed describes the same primary operator considered there. In the small $\lambda n $ regime the expansion of $\Delta_n$ may be compared to the diagrammatic result; this check was performed in \cite{Jack:2020wvs} up to six-loop order, finding perfect agreement.  In the regime $\lambda n\gg 4\pi$ the result reproduces the EFT result \eqref{eqEFTdim} with $\frac{c}{3}=\frac{4}{\sqrt{3}\lambda}$. 

It is worth recalling that the technical reason why the solution to this problem reduces to the EFT for $\lambda n\gg 1$ is that, in this regime, the solution \eqref{eqAnsatz2pt} satisfies $\rho^2\simeq 4\sqrt{3}|\pd\chi|$, while the kinetic term for the radial mode is negligible $(\pd\rho)^2\ll |\pd\chi|^2\rho^2\sim\rho^6$. We may therefore neglect derivatives of $\rho$ in the equations of motion \eqref{Eq_EOM2}, which reduce to the one deriving from \eqref{eqEFT}.

While the field should always transform as a primary with half-integer scaling dimension in $x$, the fact that the solution depends on the insertion points $x_i$ and $x_f$ as specified in Eq. \eqref{eqAnsatz2pt} is a consequence of the chosen boundary conditions. This happens because the boundary conditions \eqref{eq:chiBC} and \eqref{eq:rhoBC1} specify a primary operator \cite{Giombi:2020enj}. To appreciate this, notice that the classical expectation value for the field physically represents a (normalized) correlator of three primary operators:
\begin{equation}\label{eq_Rho_class}
\rho^2_{class.}(x)=\frac{\langle\bar{\phi}^n(x_f)\bar{\phi}\phi(x)\phi^n(x_i)\rangle}{2\langle\bar{\phi}^n(x_f)\phi^n(x_i)\rangle}\,,
\end{equation}
where we stressed with the subscript \emph{class.} that the left-hand side is evaluated on the saddle point solution. It is then easy to see that Eqs. \eqref{eqAnsatz2pt} and \eqref{eqAnsatz2pt2} ensure that Eq. \eqref{eq_Rho_class} has the right form to describe a CFT correlator of three primaries. One may similarly check that the form of the solution ensures that higher-point functions
of the form $\langle\bar{\phi}^n(x_f)\ldots\phi^n(x_i)\rangle$ are compatible with conformal invariance. Relatedly, as an obvious by-product of our analysis, we can compute an infinite number of three- and higher-point functions of light operators inserted within $\phi^n$ and $\bar{\phi}^n$.\footnote{This can also be done within the original approach of \cite{Badel}.}

\subsection{The triple-scaling limit}\label{SecTriple}

As for the two-point function \eqref{eq2pt}, the three-point function $\langle\phi^{n_1}\phi^{n_2}\bar{\phi}^{n_1+n_2}\rangle$ can be computed in the triple scaling limit $\lambda\rightarrow 0$, $n_1\rightarrow \infty$ and $n_2\rightarrow\infty$ with $\lambda n_1,\,\lambda n_2$ fixed. We assume $n_1\geq n_2$ for definiteness. The result is organized as:
\begin{multline}
\langle\bar{\phi}^{n_1+n_2}(x_f)\phi^{n_2}(x_c)\phi^{n_1}(x_i)\rangle\\
=
(n_1+n_2)!e^{\frac{1}{\lambda}G_{-1}(\lambda n_1,\lambda n_2;x_{ic},x_{cf},x_{fi})+\ldots}
\,.
\end{multline}
Comparing with the general form for the three-point function \eqref{Eq_3ptGeneral}
and taking into account the normalization of the two-point function \eqref{eq2pt},
this implies that the OPE coefficient admits an expansion of the form:
\begin{equation}\label{eqLambdaF}
\lambda_{n_1,n_2,\overline{n_1+n_2}}
=\lambda^{1/4}\exp{\left[\frac{1}{\lambda}F_{-1}(\lambda n_1,\lambda n_2)+\ldots\right]}\,.
\end{equation}

Proceeding as commented below Eq. \eqref{eq_Rho_class}, we can compute the OPE coefficient in the regime $\lambda n_2\ll 4\pi$ using the classical solution \eqref{eqAnsatz2pt} to extract the four-point function $\langle\bar{\phi}^{n_1}(x_f)\bar{\phi}^{n_2}(x_2)\phi^{n_2}(x_1)\phi^{n_1}(x_i)\rangle$ and matching the classical result to the OPE decomposition (see e.g. \cite{MoninCFT,Giombi:2020enj} for similar calculations). The result for the fusion coefficient reads:
\begin{equation}\label{OPE_small}
\lambda_{n_1,n_2,\overline{n_1+n_2}}=\frac{1}{\sqrt{n_2!}}\left(\frac{n_1}{2\mu_1}\right)^{n_2/2}
\end{equation}
where $\mu_1$ is given by Eq. \eqref{eqMu} with $n=n_1$. This is equivalent to
\begin{equation}
F_{-1}=\frac{1}{2}\lambda n_2 \left[1-\log \left(2\mu_1\frac{ \lambda n_2 }{\lambda n_1}\right)\right]+O\left(\frac{(\lambda n_2)^2}{(4\pi)^2}\right)\,.
\end{equation}
Expanding for small $\lambda n_1 $, Eq. \eqref{OPE_small} reproduces the tree-level result $\sqrt{\frac{(n_1+n_2)!}{n_1!n_2!}}$ (using Stirling's formula). In the opposite regime, the result takes the form $\sim (n_1/\lambda)^{n_2/4}$, in agreement with the EFT prediction for the OPE coefficient of a light operator of dimension $n_2/2$ and two heavy charged operators \cite{MoninCFT}. 

\subsection{The boundary value problem}\label{SecBVP}

According to the prescription explained in Sec. \ref{SecOpDefWC}, to compute the three-point function we need to solve the problem specified by the action:
\begin{equation}\label{eqAction3pt_full}
S_{eff}=S+S_B^{(i)}+S_B^{(c)}+S_B^{(f)}\,,
\end{equation}
where 
\begin{equation}
S_B^{(a)}=-i\frac{n_a}{4\pi}\int_{\pd B(x_a,r)}d\Omega\chi\,,\quad
a=i,c,f\,,
\end{equation}
with $n_i=n_1$, $n_c=n_2$ and $n_f=-n_1-n_2$. The boundary conditions for $\rho$ are specified as in Eq. \eqref{eq:rhoBC1}:
\begin{equation}
\rho^2\stackrel{x\rightarrow x_a}{\longrightarrow}\frac{f^2_a}{|x-x_a|}
\qquad\text{with}\qquad \frac{f^4_a}{48}=\mu^2_a-\frac14\,,
\end{equation}
where $\mu_a$ is given by \eqref{eqMu} with $n=n_a$.

To solve the equations of motion, it is natural to generalize the three-point function ansatz \eqref{eqAnsatz2pt} and look for solutions which take the form of a four-point function:
\begin{equation}\label{eqAnsatz3pt}
\phi\sim \langle\mathcal{O}_{\Delta=\frac12}(x)
\mathcal{O}_{\Delta=-\frac{\mu_i}{2}}(x_i)\mathcal{O}_{\Delta=-\frac{\mu_c}{2}}(x_c)\mathcal{O}_{\Delta=-\frac{\mu_{f}}{2}}(x_f)\rangle\,.
\end{equation}
Formally, this form of the solution follows from arguments analogous to those explained around Eq. \eqref{eq_Rho_class}.\footnote{From that perspective, the boundary conditions \eqref{eq:chiBC} and \eqref{eq:rhoBC1} ensure that correlation functions of light operators obey the correct OPE limits close to $x_i$, $x_c$ and $x_f$.} Working in polar field coordinates \eqref{eq_polar_coordinates}, the ansatz \eqref{eqAnsatz3pt} reads:
\begin{equation}\label{eqAnsatz3pt2}
\rho(x)=
\frac{\hat{\rho}(\tau,\theta)|x_i-x_f|^{1/2}}{|x-x_i|^{1/2}|x-x_f|^{1/2}}\,,\quad
\chi(x)=\chi(\tau,\theta)+\text{const.}\,,
\end{equation}
where the constant piece in $\chi$ is irrelevant due to the shift symmetry.  The physical meaning of this parametrization is appreciated noticing that the equations of motion for $\hat{\rho}$ and $\chi$ coincide with those for the theory on $\mathbb{R}\times S^{2}$ 
\begin{gather}\nonumber
-\nabla^\mu\pd_\mu\hat\rho+
(\pd\chi )^2\hat\rho+\frac{1}{4}\hat\rho+
\frac{1}{48}\hat\rho^5=0\,,\\ \label{eqEOMfull3}
\nabla_\mu\left(\hat\rho^2\pd^\mu\chi\right)=0\,,
\end{gather}
where $\nabla_\mu$ is the covariant derivative on $\mathbb{R}\times S^2$.
The boundary conditions set the value of the fields at $\tau\rightarrow\pm\infty$ and at the point $(\tau,\theta)=(0,0)$:
\begin{align}\nonumber
\tau\rightarrow -\infty:\;\;&\begin{cases}
\hat{\rho}\rightarrow f_i+O\left(e^{-\tau}\right) \\
\chi\rightarrow-i\mu_i\tau+\hat{c}_i+O\left(e^{\tau}\right);
\end{cases}
\\ \nonumber
(\tau,\theta)\rightarrow (0,0):\;\;&\begin{cases}
\hat{\rho}\rightarrow
f_c/v^{1/4} +O\left(v^{1/4}\right)\\
\chi\rightarrow-i\frac{\mu_c}{2}\log v
+\hat{c}_c+O\left(\sqrt{v}\right)
\,;
\end{cases}\\
\tau\rightarrow \infty:\;\;&\begin{cases}
\hat{\rho}\rightarrow f_f+O\left(e^{-\tau}\right)\\
\chi\rightarrow -i|\mu_f|\tau +\hat{c}_f+O\left(e^{-\tau}\right)\,.
\end{cases}
\label{eqBCfull3}
\end{align}
The physical interpretation of the semiclassical problem is the same as in the EFT. As below Eq. \eqref{eqBoundaryEFT}, we defined the first corrections $\hat{c}_{i/c/f}$ to the field $\chi$ close to the insertion points, whose value is determined solving the equations of motion.  

Once the classical profile has been found, the OPE coefficient can be obtained computing the classical value of the action \eqref{eqAction3pt}. Since the solution cannot develop singularities other than the ones required by the boundary conditions \eqref{eqBCfull3},\footnote{This is because, as explained below Eq. \eqref{eq_Rho_class}, the classical solution represents a physical correlator itself, whose only singularities are the ones dictated by the OPE.} all the divergent contributions are absorbed by the normalization in Eq. \eqref{eq2ptResultPre} and we are left with a finite result that can be compactly written as
\begin{equation}\label{eq3ptFull}
\lambda_{n_1,n_2,\overline{n_1+n_2}}=\exp\left[i\sum_{a}n_a\hat{c}_a+\frac{1/\lambda}{144}I_3^{(mod)}\right]\,,
\end{equation}
where $I_3^{(mod)}$ is the following convergent integral:
\begin{equation}
I_3^{(mod)}=2\pi\int_{-\infty}^\infty d\tau \int_0^\pi d\theta\sin\theta  
\hat{\rho}_{mod}^6(\tau,\theta)\,.
\end{equation}
Here $\hat{\rho}_{mod}$ is obtained subtracting the non-integrable contributions from $\hat{\rho}$:
\begin{multline}
\hat{\rho}_{mod}^6(\tau,\theta)=\hat{\rho}^6(\tau,\theta)-
\left[\frac{f_i^6+f_f^6-f_c^6}{2}\right.\\
\left.+
\frac{f_c^6+f_f^6-f_i^6}{2}
\left(\frac{u}{v}\right)^{\frac{3}{2}}
+\frac{f_i^6+f_c^6-f_f^6}{2}\frac{1}{v^{3/2}}\right]
\,.
\end{multline}
Notice that given the boundary conditions, $\chi$ is purely imaginary on the saddle point, hence so are the $\hat{c}_a$ and the OPE coefficient is real.

Eqs. \eqref{eqEOMfull3}, \eqref{eqBCfull3} and \eqref{eq3ptFull} provide an abstract general solution to the problem of finding the OPE coefficient of three large charge operators. In practice an explicit solution can only be found numerically. We leave a detailed analysis for future work. Here we just comment that for $\lambda n_1\geq \lambda n_2\gg 1$ the solution satisfies $\hat{\rho}^6\sim |\pd\chi|^3\gg(\pd\hat{\rho})^2$ close to all the insertion points. We may hence treat self-consistently the kinetic term of the radial mode as a perturbation, and to leading order the problem reduces to the EFT one studied in Sec. \ref{SecNumerics}. 

\bibliography{Biblio}

\providecommand{\href}[2]{#2}\begingroup\raggedright\begin{thebibliography}{10}

\bibitem{landau1958quantum}
L.~D. Landau and E.~M. Lifshitz, \emph{Quantum Mechanics: Non-relativistic
  Theory. V. 3 of Course of Theoretical Physics}. Pergamon Press, 1958.

\bibitem{MoninCFT}
A.~Monin, D.~Pirtskhalava, R.~Rattazzi and F.~K. Seibold, \emph{{Semiclassics,
  Goldstone Bosons and CFT data}},
  \href{https://doi.org/10.1007/JHEP06(2017)011}{\emph{JHEP} {\bfseries 06}
  (2017) 011} [\href{https://arxiv.org/abs/1611.02912}{{\ttfamily
  1611.02912}}].

\bibitem{Berenstein:2002jq}
D.~E. Berenstein, J.~M. Maldacena and H.~S. Nastase, \emph{{Strings in flat
  space and pp waves from N=4 superYang-Mills}},
  \href{https://doi.org/10.1088/1126-6708/2002/04/013}{\emph{JHEP} {\bfseries
  04} (2002) 013} [\href{https://arxiv.org/abs/hep-th/0202021}{{\ttfamily
  hep-th/0202021}}].

\bibitem{Alday:2007mf}
L.~F. Alday and J.~M. Maldacena, \emph{{Comments on operators with large
  spin}}, \href{https://doi.org/10.1088/1126-6708/2007/11/019}{\emph{JHEP}
  {\bfseries 11} (2007) 019} [\href{https://arxiv.org/abs/0708.0672}{{\ttfamily
  0708.0672}}].

\bibitem{Komargodski:2012ek}
Z.~Komargodski and A.~Zhiboedov, \emph{{Convexity and Liberation at Large
  Spin}}, \href{https://doi.org/10.1007/JHEP11(2013)140}{\emph{JHEP} {\bfseries
  11} (2013) 140} [\href{https://arxiv.org/abs/1212.4103}{{\ttfamily
  1212.4103}}].

\bibitem{Fitzpatrick:2012yx}
A.~L. Fitzpatrick, J.~Kaplan, D.~Poland and D.~Simmons-Duffin, \emph{{The
  Analytic Bootstrap and AdS Superhorizon Locality}},
  \href{https://doi.org/10.1007/JHEP12(2013)004}{\emph{JHEP} {\bfseries 12}
  (2013) 004} [\href{https://arxiv.org/abs/1212.3616}{{\ttfamily 1212.3616}}].

\bibitem{Hellerman:2013kba}
S.~Hellerman and I.~Swanson, \emph{{String Theory of the Regge Intercept}},
  \href{https://doi.org/10.1103/PhysRevLett.114.111601}{\emph{Phys. Rev. Lett.}
  {\bfseries 114} (2015) 111601}
  [\href{https://arxiv.org/abs/1312.0999}{{\ttfamily 1312.0999}}].

\bibitem{Hellerman}
S.~Hellerman, D.~Orlando, S.~Reffert and M.~Watanabe, \emph{{On the CFT
  Operator Spectrum at Large Global Charge}},
  \href{https://doi.org/10.1007/JHEP12(2015)071}{\emph{JHEP} {\bfseries 12}
  (2015) 071} [\href{https://arxiv.org/abs/1505.01537}{{\ttfamily
  1505.01537}}].

\bibitem{Son:2002zn}
D.~T. Son, \emph{{Low-energy quantum effective action for relativistic
  superfluids}},  \href{https://arxiv.org/abs/hep-ph/0204199}{{\ttfamily
  hep-ph/0204199}}.

\bibitem{Hellerman:2017veg}
S.~Hellerman, S.~Maeda and M.~Watanabe, \emph{{Operator Dimensions from
  Moduli}}, \href{https://doi.org/10.1007/JHEP10(2017)089}{\emph{JHEP}
  {\bfseries 10} (2017) 089}
  [\href{https://arxiv.org/abs/1706.05743}{{\ttfamily 1706.05743}}].

\bibitem{Hellerman:2017sur}
S.~Hellerman and S.~Maeda, \emph{{On the Large $R$-charge Expansion in
  ${\mathcal N} = 2$ Superconformal Field Theories}},
  \href{https://doi.org/10.1007/JHEP12(2017)135}{\emph{JHEP} {\bfseries 12}
  (2017) 135} [\href{https://arxiv.org/abs/1710.07336}{{\ttfamily
  1710.07336}}].

\bibitem{Cuomo:2017vzg}
G.~Cuomo, A.~de~la Fuente, A.~Monin, D.~Pirtskhalava and R.~Rattazzi,
  \emph{{Rotating superfluids and spinning charged operators in conformal field
  theory}}, \href{https://doi.org/10.1103/PhysRevD.97.045012}{\emph{Phys. Rev.
  D} {\bfseries 97} (2018) 045012}
  [\href{https://arxiv.org/abs/1711.02108}{{\ttfamily 1711.02108}}].

\bibitem{Badel}
G.~Badel, G.~Cuomo, A.~Monin and R.~Rattazzi, \emph{{The Epsilon Expansion
  Meets Semiclassics}},
  \href{https://doi.org/10.1007/JHEP11(2019)110}{\emph{JHEP} {\bfseries 11}
  (2019) 110} [\href{https://arxiv.org/abs/1909.01269}{{\ttfamily
  1909.01269}}].

\bibitem{Jafferis:2017zna}
D.~Jafferis, B.~Mukhametzhanov and A.~Zhiboedov, \emph{{Conformal Bootstrap At
  Large Charge}}, \href{https://doi.org/10.1007/JHEP05(2018)043}{\emph{JHEP}
  {\bfseries 05} (2018) 043}
  [\href{https://arxiv.org/abs/1710.11161}{{\ttfamily 1710.11161}}].

\bibitem{Grassi:2019txd}
A.~Grassi, Z.~Komargodski and L.~Tizzano, \emph{{Extremal Correlators and
  Random Matrix Theory}},  \href{https://arxiv.org/abs/1908.10306}{{\ttfamily
  1908.10306}}.

\bibitem{Bargheer:2019kxb}
T.~Bargheer, F.~Coronado and P.~Vieira, \emph{{Octagons I: Combinatorics and
  Non-Planar Resummations}},
  \href{https://doi.org/10.1007/JHEP08(2019)162}{\emph{JHEP} {\bfseries 08}
  (2019) 162} [\href{https://arxiv.org/abs/1904.00965}{{\ttfamily
  1904.00965}}].

\bibitem{Bargheer:2019exp}
T.~Bargheer, F.~Coronado and P.~Vieira, \emph{{Octagons II: Strong Coupling}},
  \href{https://arxiv.org/abs/1909.04077}{{\ttfamily 1909.04077}}.

\bibitem{Hellerman:2020sqj}
S.~Hellerman, S.~Maeda, D.~Orlando, S.~Reffert and M.~Watanabe,
  \emph{{S-duality and correlation functions at large R-charge}},
  \href{https://arxiv.org/abs/2005.03021}{{\ttfamily 2005.03021}}.

\bibitem{Lashkari:2016vgj}
N.~Lashkari, A.~Dymarsky and H.~Liu, \emph{{Eigenstate Thermalization
  Hypothesis in Conformal Field Theory}},
  \href{https://doi.org/10.1088/1742-5468/aab020}{\emph{J. Stat. Mech.}
  {\bfseries 1803} (2018) 033101}
  [\href{https://arxiv.org/abs/1610.00302}{{\ttfamily 1610.00302}}].

\bibitem{Banerjee:2017fcx}
D.~Banerjee, S.~Chandrasekharan and D.~Orlando, \emph{{Conformal dimensions via
  large charge expansion}},
  \href{https://doi.org/10.1103/PhysRevLett.120.061603}{\emph{Phys. Rev. Lett.}
  {\bfseries 120} (2018) 061603}
  [\href{https://arxiv.org/abs/1707.00711}{{\ttfamily 1707.00711}}].

\bibitem{Gross:1987ar}
D.~J. Gross and P.~F. Mende, \emph{{String Theory Beyond the Planck Scale}},
  \href{https://doi.org/10.1016/0550-3213(88)90390-2}{\emph{Nucl. Phys. B}
  {\bfseries 303} (1988) 407}.

\bibitem{Cardy:2017qhl}
J.~Cardy, A.~Maloney and H.~Maxfield, \emph{{A new handle on three-point
  coefficients: OPE asymptotics from genus two modular invariance}},
  \href{https://doi.org/10.1007/JHEP10(2017)136}{\emph{JHEP} {\bfseries 10}
  (2017) 136} [\href{https://arxiv.org/abs/1705.05855}{{\ttfamily
  1705.05855}}].

\bibitem{Collier:2019weq}
S.~Collier, A.~Maloney, H.~Maxfield and I.~Tsiares, \emph{{Universal dynamics
  of heavy operators in CFT$_{2}$}},
  \href{https://doi.org/10.1007/JHEP07(2020)074}{\emph{JHEP} {\bfseries 07}
  (2020) 074} [\href{https://arxiv.org/abs/1912.00222}{{\ttfamily
  1912.00222}}].

\bibitem{Belin:2020hea}
A.~Belin and J.~de~Boer, \emph{{Random Statistics of OPE Coefficients and
  Euclidean Wormholes}},  \href{https://arxiv.org/abs/2006.05499}{{\ttfamily
  2006.05499}}.

\bibitem{Simmons-Duffin:2016gjk}
D.~Simmons-Duffin, \emph{{The Conformal Bootstrap}},  in \emph{{Theoretical
  Advanced Study Institute in Elementary Particle Physics}: {New Frontiers in
  Fields and Strings}}, pp.~1--74, 2017,
  \href{https://arxiv.org/abs/1602.07982}{{\ttfamily 1602.07982}},
  \href{https://doi.org/10.1142/9789813149441_0001}{DOI}.

\bibitem{Krikun:2018ufr}
A.~Krikun, \emph{{Numerical Solution of the Boundary Value Problems for Partial
  Differential Equations. Crash course for holographer}},  1, 2018,
  \href{https://arxiv.org/abs/1801.01483}{{\ttfamily 1801.01483}}.

\bibitem{Cuomo:2020rgt}
G.~Cuomo, \emph{{A note on the large charge expansion in 4d CFT}},
  \href{https://doi.org/10.1016/j.physletb.2020.136014}{\emph{Phys. Lett. B}
  {\bfseries 812} (2021) 136014}
  [\href{https://arxiv.org/abs/2010.00407}{{\ttfamily 2010.00407}}].

\bibitem{Alvarez-Gaume:2019biu}
L.~Alvarez-Gaume, D.~Orlando and S.~Reffert, \emph{{Large charge at large N}},
  \href{https://doi.org/10.1007/JHEP12(2019)142}{\emph{JHEP} {\bfseries 12}
  (2019) 142} [\href{https://arxiv.org/abs/1909.02571}{{\ttfamily
  1909.02571}}].

\bibitem{Giombi:2020enj}
S.~Giombi and J.~Hyman, \emph{{On the Large Charge Sector in the Critical
  $O(N)$ Model at Large $N$}},
  \href{https://arxiv.org/abs/2011.11622}{{\ttfamily 2011.11622}}.

\bibitem{Delacretaz:2020nit}
L.~V. Delacretaz, \emph{{Heavy Operators and Hydrodynamic Tails}},
  \href{https://doi.org/10.21468/SciPostPhys.9.3.034}{\emph{SciPost Phys.}
  {\bfseries 9} (2020) 034} [\href{https://arxiv.org/abs/2006.01139}{{\ttfamily
  2006.01139}}].

\bibitem{Rubakov:1995hq}
V.~A. Rubakov, \emph{{Nonperturbative aspects of multiparticle production}},
  in \emph{{2nd Rencontres du Vietnam}: {Consisting of 2 parallel conferences:
  Astrophysics Meeting: From the Sun and Beyond / Particle Physics Meeting:
  Physics at the Frontiers of the Standard Model}}, 10, 1995,
  \href{https://arxiv.org/abs/hep-ph/9511236}{{\ttfamily hep-ph/9511236}}.

\bibitem{Son:1995wz}
D.~T. Son, \emph{{Semiclassical approach for multiparticle production in scalar
  theories}}, \href{https://doi.org/10.1016/0550-3213(96)00386-0}{\emph{Nucl.
  Phys. B} {\bfseries 477} (1996) 378}
  [\href{https://arxiv.org/abs/hep-ph/9505338}{{\ttfamily hep-ph/9505338}}].

\bibitem{Khoze:2018mey}
V.~V. Khoze and J.~Reiness, \emph{{Review of the semiclassical formalism for
  multiparticle production at high energies}},
  \href{https://doi.org/10.1016/j.physrep.2019.06.004}{\emph{Phys. Rept. C}
  {\bfseries 822} (2019) 1} [\href{https://arxiv.org/abs/1810.01722}{{\ttfamily
  1810.01722}}].

\bibitem{Monin:2018cbi}
A.~Monin, \emph{{Inconsistencies of higgsplosion}},
  \href{https://arxiv.org/abs/1808.05810}{{\ttfamily 1808.05810}}.

\bibitem{Badel2}
G.~Badel, G.~Cuomo, A.~Monin and R.~Rattazzi, \emph{{Feynman diagrams and the
  large charge expansion in $3-\varepsilon$ dimensions}},
  \href{https://doi.org/10.1016/j.physletb.2020.135202}{\emph{Phys. Lett. B}
  {\bfseries 802} (2020) 135202}
  [\href{https://arxiv.org/abs/1911.08505}{{\ttfamily 1911.08505}}].

\bibitem{Antipin:2020abu}
O.~Antipin, J.~Bersini, F.~Sannino, Z.-W. Wang and C.~Zhang, \emph{{Charging
  the $O(N)$ model}},
  \href{https://doi.org/10.1103/PhysRevD.102.045011}{\emph{Phys. Rev. D}
  {\bfseries 102} (2020) 045011}
  [\href{https://arxiv.org/abs/2003.13121}{{\ttfamily 2003.13121}}].

\bibitem{Pisarski}
R.~D. Pisarski, \emph{Fixed-point structure of
  ${({\ensuremath{\phi}}^{6})}_{3}$ at large $n$},
  \href{https://doi.org/10.1103/PhysRevLett.48.574}{\emph{Phys. Rev. Lett.}
  {\bfseries 48} (1982) 574}.

\bibitem{Arias-Tamargo:2019kfr}
G.~Arias-Tamargo, D.~Rodriguez-Gomez and J.~G. Russo, \emph{{Correlation
  functions in scalar field theory at large charge}},
  \href{https://doi.org/10.1007/JHEP01(2020)171}{\emph{JHEP} {\bfseries 01}
  (2020) 171} [\href{https://arxiv.org/abs/1912.01623}{{\ttfamily
  1912.01623}}].

\bibitem{Jack:2020wvs}
I.~Jack and D.~R.~T. Jones, \emph{{Anomalous dimensions for $\phi^n$ in scale
  invariant $d=3$ theory}},
  \href{https://doi.org/10.1103/PhysRevD.102.085012}{\emph{Phys. Rev. D}
  {\bfseries 102} (2020) 085012}
  [\href{https://arxiv.org/abs/2007.07190}{{\ttfamily 2007.07190}}].

\end{thebibliography}\endgroup
	\bibliographystyle{JHEP.bst}

\end{document}